\documentclass[mypaper,8pt,twoside]{CoAst}
\usepackage{epsf,graphicx,fancyhdr,sfmath}
\renewcommand{\chaptermark}[1]{\markboth{#1}{}}

\newcommand{\apj}{ApJ}
\newcommand{\aap}{A\&A}  
\newcommand{\mnras}{MNRAS}

\newcommand{\teff}{\ensuremath{T_{eff}}}             
\newcommand{\logg}{\ensuremath{\log g}}                     

\newcommand{\kopf}{\small\itshape Comm.\ in Asteroseismology, N$^{\textsf{\underline{o}}}$ 159, 2009\\
Proceedings of the JENAM 2008 Symposium N$^{\textsf{\underline{o}}}$~4:
Asteroseismology and Stellar Evolution}

\newcommand{\Authors}[1]{\begin{center}\normalsize\bf\sf #1 \end{center}}

\renewcommand{\author}[1]{\begin{center}\normalsize\bf\sf #1 \end{center}}
\newcommand{\Address}[1]{\begin{center}\small\sf #1 \end{center}}

\newcommand{\Objects}[1]{{\vspace{3mm}\small \noindent Individual Objects: }\small\sf \hangindent=27truemm \hangafter=1 #1 \normalsize}

\renewenvironment{abstract}{\section*{Abstract}\normalsize\sf}{}
\newcommand{\References}[1]{\begin{flushleft}{\large References\\}\vspace*{2mm}\small #1 \end{flushleft}}

\newcommand{\chapterCoAst}[2]{\chapter[\sf\normalsize #1\\ \footnotesize \hspace*{5mm}by #2 \sf\normalsize][]{#1\\}\rhead[\fancyplain{}{\sf\footnotesize \center{#1}}]{\fancyplain{}{\sffamily\thepage}}\lhead[\fancyplain{\kopf}{\sffamily\thepage}]{\fancyplain{\kopf}{\sf\footnotesize \center{#2}}}}

\newcommand{\figureCoAst}[5]{\begin{figure}[#4]
\centering
\includegraphics*[#5]{#1}
\caption{#2}
\label{#3}
\end{figure}}

\renewcommand{\chaptername}{}

\newcommand{\acknowledgments}[1]{\vspace*{5mm}\noindent  \textbf{Acknowledgments.} #1}

\newcommand{\figureCoAstII}[6]{\begin{figure}[#5]
\centering
\includegraphics*[#6]{#1}
\includegraphics*[#6]{#2}
\caption{#3}
\label{#4}
\end{figure}}

\def\rfr{\smallskip\par\noindent
        \hangindent=7truemm
        \hangafter=1}
\def\sol{\ensuremath{_\odot}}

\begin{document}
\sf

\chapterCoAst{Asteroseismology and evolution of EHB stars}{R.H.\,\O stensen}
\Authors{R.\,H.\,\O stensen} 
\Address{
  Instituut voor Sterrenkunde, K.U.Leuven,
  Celestijnenlaan 200D, B-3001 Leuven, Belgium
}

\noindent
\begin{abstract}
The properties of the Extreme Horizontal Branch stars are quite well
understood, but much uncertainty surrounds the many paths that bring a
star to this peculiar configuration.
Asteroseismology of pulsating EHB stars has been performed on a
number of objects, bringing us to the stage where comparisons of the
inferred properties with evolutionary models becomes feasible.
In this review I will outline our current understanding of the formation
and evolution of these stars, with emphasis on recent progress.
The aim is to show how the physical parameters derived by
asteroseismology can enable the discrimination
between different evolutionary models.
\end{abstract}

\Objects{V361 Hya, V1093 Her, DW Lyn, V391 Peg, Balloon 090100001, KL UMa, NY Vir, V338 Ser, LS\,IV--14$^\circ$116, HD 188112}

\section*{Introduction}
Let me first clarify the basic terminology with respect to EHB stars, which
can sometimes be confusing as the terms EHB and sdB are often used to label
the same stars. The terms Extended Horizontal Branch or Extreme Horizontal
Branch have been used interchangeably to describe the sequence of stars
observed to lie bluewards of the normal Horizontal Branch stars in
globular clusters, and also in temperature/gravity plots of hot field stars.
The EHB feature was first described and associated with field sdB and
sdO stars by Greenstein \& Sargent (1974).
Now, EHB stars are taken to mean core helium burning stars with an
envelope too thin to sustain hydrogen burning. It is also understood
that not all sdB stars are EHB stars. In particular, if a star loses
its envelope without the core reaching the mass required for the helium
flash, its cooling track can take it through the sdB domain
on its way to become a helium core WD. The sdB/sdO terms are used 
to describe the spectroscopic appearance and do not presume any
particular evolutionary stage. Several subclassification schemes have been
used, but most common nowadays is the one introduced by
Moehler et al.~(1990).
This scheme names as sdB stars those of the hot subdwarfs showing
He\,I absorption lines, as sdO stars those showing He\,II, and as sdOB
stars those showing features of both.
Additionally the terms He-sdB and He-sdO are used to describe stars
in which the helium lines dominate over the Balmer lines.
The EHB forms a sequence of stars from the coolest sdBs to
the sdOB domain, and it is clear that most stars given this
classification are in fact EHB stars. For the He-rich objects
a coherent picture has yet to emerge.

The current canonical picture of the EHB stars was mostly established
by Heber (1986), in which the EHB stars are helium core
burning stars with masses close to the core helium flash mass
of $\sim$0.47\,M\sol, and an extremely thin hydrogen envelope, too
thin to sustain hydrogen burning (no more than 1\% by mass).
It is understood that they are post red giant branch (RGB) stars
that have started
core helium burning in a helium flash before or after the envelope
was removed by any of several possible mechanisms.
The lifetime of EHB stars from the zero-age EHB (ZAEHB) to the
terminal age EHB (TAEHB), when core helium runs out, takes between
100 and 150\,Myrs.
The post-EHB evolution will take them through the sdO
domain directly to the white dwarf (WD) cooling curve without ever
passing through a second giant stage.
The time they spend shell helium burning before
leaving the sdO domain can be up to 20\,Myrs.

Although the future evolution of EHB stars after core He-exhaustion has
always been presumed quite simple, the paths that
lead to the EHB have always been somewhat mysterious.
New hope that the evolutionary paths leading to the formation of
EHB stars can be resolved has been kindled by the discovery that many
of them pulsate, which has opened up the possibility of
probing their interiors using asteroseismological methods.
These pulsators are known as sdBV stars, and several distinct subclasses
are now recognised (see the {\em Asteroseismology} section below).
But in order to understand what questions asteroseismology can ask and
answer, it is essential to understand the different paths
that produce EHB stars. Only by understanding the evolutionary history of
these stars is it possible to construct realistic models of their
interiors which are needed for asteroseismology to be able to distinguish
between the different formation scenarios.
For this reason we will review the essential points of
the {\em Formation and Evolution} first, after starting with a look
at the observed properties
of the hot subdwarf population in {\em The Observed EHB} below. 

Besides the spectacularly rapid pulsations in the EHB stars,
another factor that has contributed to the recent burst in interest in
EHB stars is the realisation that these stars are the main contributor
to the UV-upturn phenomenon observed in elliptical galaxies.
An excellent review of the UV upturn and the binary population synthesis
models required to model this phenomenon, can be found in
Podsiadlowski et al.~(2008).
For a more in-depth review of the properties of all hot subdwarf stars,
the exhaustive review by Heber (2009) is recommended.

\section*{The Observed EHB}

Hot subdwarf stars were found in the galactic caps already by
Humason \& Zwicky (1947). By the time Greenstein \& Sargent (1974)
wrote their seminal paper, the number of such faint blue stars had grown 
to 189, permitting a systemic study of the population.
The PG survey (Green et al.~1986), which covered more than 10\,000 square
degrees at high galactic latitudes, found that of 1874 UV-excess objects
detected more than 1000 were hot subdwarfs, so these stars dominate the
population of faint blue stars down to the PG survey limit ($B$\,=\,16.5).
Together with the large sample of subdwarfs detected in the HS survey 
and analysed by Edelmann et al.~(2003),
these have provided a rich source of hot subdwarfs
for observers to follow up, and new discoveries are still being made.
The recent Sloan Digital Sky Survey (SDSS, Stoughton et al.~2002),
also contains spectra of more than 1000 hot subdwarfs, but as 
the SDSS reaches much deeper than the PG survey, WD stars start to
dominate around about $B$\,=\,18 as the thickness of the galactic disk
is reached. The Subdwarf Database (\O stensen 2006) catalogs about 2500
hot subdwarfs, with extensive references to the available literature.

\figureCoAst{BG_TGHe+EHB}
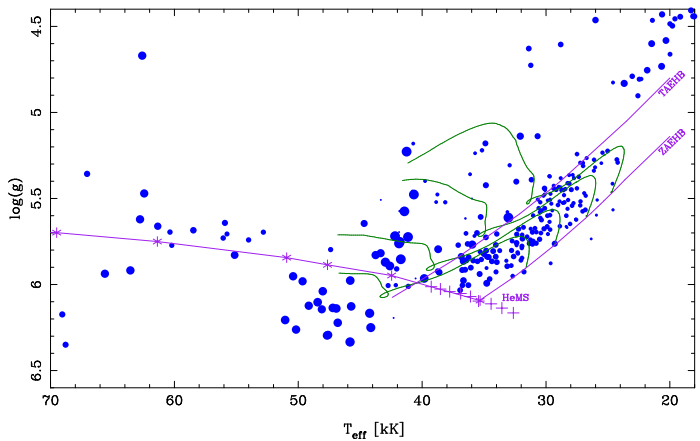
{The EHB in the \teff--\logg\ plane as observed by the Bok-Green survey
(Green et al.~2008).  The symbols mark observed stars with the
size indicating the helium abundance.
The theoretical zero age HeMS is shown for a wide range of masses.
Models from Paczy\'nski (1971) with masses of 0.5, 0.7, 0.85, 1, 1.5
and 2\,M\sol\ are marked with $\ast$ symbols (starting from low \teff). 
More recent models from Kawaler \& Hostler (2005) are shown
for M$_\ast$ = 0.41, 0.43, ... 0.57 M\sol\ are
marked with + symbols, and the zero age and terminal age EHB
for the 0.47\,M\sol\ core models are also drawn.
For the latter, four evolutionary tracks with different envelope
thicknesses $\log M_e/M_\ast$\,=\,-3.5,-3,-2.5,-2 are drawn
(starting from high \logg ).
}
{fig:TG_EHB}{t}{clip,angle=0,width=0.8\textwidth}

Several surveys have attempted to tackle the question of the binary
frequency on the EHB, but the matter is complicated by the many different
types of systems EHB stars are found in. EHB stars with FGK companions
are easily detected from their double-lined spectra or from IR excess.
But EHB stars with WD or M-dwarf companions show no such features.
When the orbital periods are sufficiently short, these systems can
easily be revealed from their RV variations.
Using the RV method, Maxted et al.~(2001)
targeted 36 EHB stars and found 21 binaries, all with periods less than
30 days. This gave a fraction of short period binaries of 60\,$\pm$\,8\,\%.
Other surveys have found smaller fractions, but they have not constrained
the sample to focus strictly on the EHB.
From high-resolution VLT spectra of 76 sdB stars from the SPY survey
Lisker et al.~(2005) found that 24 
showed the signature of an FGK companion, none of which show any
RV variability. Napiwotzki et al.~(2005) reported that of 46 sdB stars in
the same sample, 18 (39\%) were RV variable. Clearly, the binary fraction
in EHB stars is much higher than for normal stars,
but an accurate number has yet to be established.

Most recently, Green et al.~(2008) have presented a uniform high
signal-to-noise low-resolution survey of a large sample including most
known hot subdwarf stars brighter than $V$\,=\,14.2, using the university
of Arizona 2.3\,m Bok telescope (hereafter referred to as the
Bok-Green or BG survey).
From this large sample the clearest
picture of the EHB to date emerges (Fig.\,1 and 2).
Most stars in the diagram are clearly well bound by EHB models
for a narrow mass distribution.
Most of the remaining stars are consistent with post-EHB models, but could
also fit core helium burning objects with higher than canonical masses.
The most helium rich objects, however, appear to form their
own sequence, which cannot be explained by canonical EHB models.
The tail of the regular horizontal branch reaches into the
diagram from the upper right, but is separated from the EHB by a
substantial gap, as first noted by Newell (1973).
Although the details of this survey are still under analysis, several
new features have been noted. The sequence of He-rich objects around
40\,kK is not compatible with current evolutionary scenarios,
since post-EHB and post-RGB objects pass too rapidly through this region
of the \teff--\logg\ plane to produce the observed clustering, but
the {\em late hot flasher} scenario (discussed in the next section)
holds some promise.

The large
group of stars below the helium main sequence (HeMS) is more problematic
as no type of star should be able to stay in this position in the
\teff--\logg\ plane more than briefly, and no clustering should occur.
The feature was also noted by Stroeer et al.~(2007), but it cannot
be ruled out that this is an artifact of the models, as many of these
stars appear to have significant amounts of CNO processed material in
their atmospheres, and the NLTE models used do not account for this.
Another remarkable feature appears when looking at the distribution
of short period binaries in the \teff--\logg\ plane (Fig.\,2). 
In particular, the incidence
of such binaries appears to be much smaller at the hot, high
gravity end of the EHB strip than among the cooler subdwarfs. This
can be understood in terms of the relative efficiency of the common
envelope ejection channel (CE, see next section)
that produces short period binaries versus the other channels
producing long period systems or single stars.
It would appear that the CE channel is significantly less effective
in removing the envelope than the other channels.

\figureCoAst{BG_TGRV}
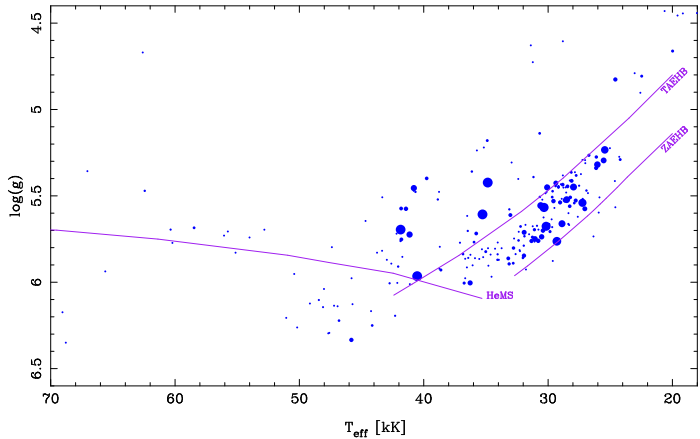
{The same as Fig.\,1, but with the symbol size indicating the dispersion
in radial velocity. The EHB stars with the highest velocity
variations appear to be concentrated at lower gravities on the EHB.
sdB+FGK stars are not
included here, due to difficulties in reliably disentangling such
composite spectra.}
{fig:TG_RV}{t}{clip,angle=0,width=0.8\textwidth}

\section*{Formation and evolution}

As binary interactions are a key for understanding the formation of
EHB stars, I will attempt a short introduction here.
Mass transfer during close binary evolution is well understood,
although there are still some unknown factors.  While the full picture is
exceedingly complex, and would take far too much time to describe here,
I will try to outline some of the most important possibilities.

\subsection*{Losing the Envelope}

There are two fundamental evolutionary paths, and which one a system enters
depends only on the mass ratio of the system.
If the expanding mass donor is more massive than the accretor, the orbit will
shrink catastrophically and the system enters a common envelope (CE) phase.
As the orbit shrinks further due to friction, orbital energy is deposited
in the envelope, spinning it up. When sufficient energy is deposited
the envelope is ejected.  Stars with an initial main sequence (MS)
mass below about 1.8\,M\sol\ can ignite helium in a core flash
before the tip of the RGB, and if the envelope is ejected at the
right time the result is an EHB star with a mass close to
the flash mass of $\sim$0.47\,M\sol, and a very close low mass companion.
If the envelope is ejected before the core reaches the required mass, the
core never ignites helium and the star will not settle on the EHB, but
continues to contract and ends up as a helium core white dwarf.
The close sdB+WD binary HD\,188112 (Heber et al.~2003) is
a particular point case, as the Hipparcos parallax together with the observed
spectroscopic surface gravity clearly constrain the mass of the subdwarf
component to be below the core helium burning limit.
If the MS mass is higher than about 2.0\,M\sol\ the
star will ignite helium non-degeneratively well before the core reaches
the mass required for the helium flash.
If the envelope is subsequently ejected on the tip of the RGB the
outcome would be an EHB star which could have a mass as low as 0.33\,M\sol.

On the other hand, if the companion is more massive than the red giant
donor filling its Roche lobe, the orbit expands and no CE
is formed. In this stable Roche lobe overflow (RLOF) scenario
the orbital period can end up as long as 2000 days.
As with CE ejection, if the red giant starts out with a mass
below the mass required for the helium flash to occur,
the mass transfer must happen close to the tip of the RGB,
and the core of the giant becomes an EHB star with a mass
close to 0.47\,M\sol.
If it is too massive, non-degenerate helium burning
starts earlier, and the result is an sdB star with a mass between 0.33 and
1.1\,M\sol\ (Han et al.~2000).

\subsection*{The Problematic Singles}

A significant number of sdB stars are definitely single stars, and their
formation is the most problematic and controversial.
While post-CE systems leave behind a close binary that is easily detectable
from the radial velocity (RV) variations, post-RLOF binaries have such
long periods that it requires very long term efforts with
high precision spectroscopy to detect them. Up to now there are no
detections of any such orbits. However, long term asteroseismic
monitoring can detect orbitally induced variations in the pulsation period 
with much higher precision than can be done from spectroscopy
(Silvotti, these proceedings). The clearest case yet is V391\,Peg where
the modulation of the pulsations are consistent with a planet with a mass
($M\sin i$) of 3.2\,$M_{\mathrm{Jupiter}}$
in an orbit with period of 1\,170 days (Silvotti et al.~2007). While the
planet might have entered the outer layers of the envelope of the
red giant before the envelope was lost, the current orbit is too wide
for it to have been responsible for the actual envelope ejection.

Several scenarios have been proposed that may produce single EHB stars.
It is well known that RGB stars lose significant mass in the form of a
stellar wind as they expand and their surface gravity becomes extremely
low.  D'Cruz et al.~(1996) computed evolutionary models of RGB stars with
mass loss parameterised by the Reimers efficiency $\eta_R$.
They found that the observed distributions of HB and EHB stars can
be explained ``{\em so long as nature provides a broad enough distribution
in $\eta_R$}''. However, the actual physics behind the large variation
in the mass loss efficiency remains unexplained.
Another possibility is the merger of two He-core WD stars, first proposed
by Webbink (1984). Saio \& Jeffery (2000) have shown that models for
such a merger can successfully predict the behaviour of the pulsating
helium star V652\,Her, demonstrating the feasibility of the merger scenario.
Such extreme helium stars will evolve to become hot helium rich subdwarfs
located close to the HeMS. However, a remaining problem
with merger models is that they invariably leave behind rapidly rotating
objects. So far, no single hot subdwarf has demonstrated more than
moderate rotational velocities from high resolution spectroscopy.

Another possibility is that CE ejection can be triggered by a
giant planet that evaporates in the process (Soker 1998). 
Nelemans \& Tauris (1998) demonstrated in the context of white dwarf
evolution that there are clear domains in initial orbital
period versus planetary mass, where the planet ejects the envelope
and is disrupted as it fills its own Roche-lobe after the spiral in.
The final rotation period of the remaining helium core is not affected
by this process, as the planet transfers almost all of its angular
momentum to the envelope before its ejection.
A final possibility for the formation of single hot subdwarfs, also
noted by Nelemans \& Tauris (1998) in the context of formation
of undermassive white dwarfs, is a variation of the RLOF mechanism.
If the envelope is transferred onto an accretor that is already
a massive white dwarf, an asymmetric accretion induced collapse may
produce a high velocity neutron star which escapes the system.
If the companion is sufficiently massive to explode as a supernova,
Marietta et al.~(2000) have computed that the explosion itself impacts
1000 times more energy on the envelope than its binding energy, easily
stripping the giant to the core. If the core is massive enough for
helium burning, and the SN explosion sufficiently asymmetric,
the remnant of the mass donor would end up as a single EHB star.
In both cases the disruption of such a binary system would leave
the EHB star with an unusual galactic orbit, which should be observable
at least in a sufficiently large sample.

\subsection*{To Flash or Not to Flash}

If an RGB star loses its envelope before the core has reached
the mass required for the helium flash, the core will contract
and heat up, before cooling as a helium core WD. The tracks
calculated for such flashless post-RGB evolution covers a wide span in 
temperatures, with models around 0.2\,M\sol\ passing
through the cool end of the EHB, and the remnants with masses
close to the helium flash mass reaching temperatures up to 100\,kK
(Driebe et al.~1999).

A borderline case exists when the mass is just on the
limit for the helium core flash to occur. Then the flash can
happen after the RGB stage, while the core is either on its way to or
on the actual WD cooling curve. Such models are known as hot flashers,
and the eventual outcome depends on the exact stage at which the
helium flash occurs.  If ignition occurs before the turning point on the
WD cooling track (early hot flashers), the remaining H-burning shell
produces a sufficient entropy barrier to prevent the convection zone produced
by the helium flash to reach the surface (Iben 1976).
But if helium ignites on the actual WD cooling curve, any
remaining shell H-burning is too weak to prevent the convection
zone reaching the envelope.  The envelope hydrogen is then
mixed into the core and quickly burnt (Sweigart 1997), and
CNO processed material is transported to the outer layers in a flash
mixing process. Such {\em late hot flashers} are predicted to end up
with an atmosphere almost totally devoid of hydrogen and with observable
CNO lines in their spectra.

Recently, Miller Bertolami et al.~(2008) have performed extensive simulations
of late hot flashers in order to determine how well models can reproduce
observations. They predict that the core flash cycles should take
place in the region above the HeMS where the strip of helium rich subdwarfs
are concentrated (Fig.\,1). However, the core flash phase lasts less than 2
Myr, after which the stars settle close to the HeMS,
for a regular EHB lifetime of at least 66\,Myr.
Observations do not support such a concentration of objects at this
location in the \teff--\logg\ plane.
To resolve this they propose that some remaining hydrogen could have survived
the mixing and should slowly diffuse to the surface, effectively
pulling the star up toward the cooler region of the EHB.

Another very recent development was presented by Politano et al.~(2008).
They have extended the classic common envelope ejection mechanism to include
the case when a low mass MS star or brown dwarf merges with the helium core,
in order to produce single EHB stars. However, as with the helium
white dwarf merger scenario, the problem remains that the products
end up spinning close to break-up velocity.
Since a rapidly rotating subpopulation
of sdB stars has yet to be found, this channel is only of marginal interest,
unless a way to eject the envelope without spinning up the core can be found.

\figureCoAst{BG_TGHe+sdBV}
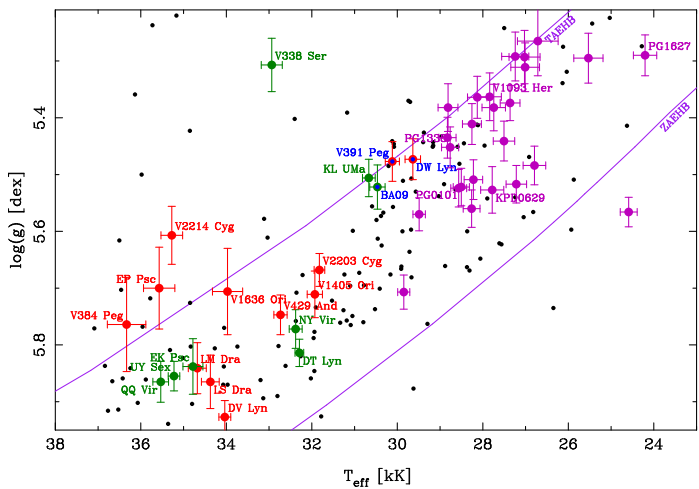
{Section of the \teff--\logg\ plane where the EHB stars are located.
Pulsators with temperatures and gravities in the BG survey are marked
with big symbols and error bars. Small symbols without error bars are
stars not observed to pulsate.
In the online version the colors indicate
short period pulsators with ({\em green}) and without ({\em red}\/)
published asteroseismic solution, long period pulsators ({\em magenta})
and hybrid pulsators ({\em blue core}).
}
{fig:TG_sdBV}{t}{clip,angle=0,width=0.8\textwidth}

\section*{Asteroseismology}

The first evidence of rapid pulsations in EHB stars were reported by
Kilkenny et al.~(1997), after their detection of multiperiodic pulsations in 
V361\,Hya.
The V361\,Hya stars span the hot end of the EHB strip
from about 28 to 34\,kK, and pulsate in $p$-modes of low 
$\ell$ orders and with photometric amplitudes up to 6\,\%.
The pulsation periods range between 100 and 400\,s, and 40 such
stars are known in the literature to date (Oreiro et al.~2009).
One of these, V338\,Ser,
has periods reaching almost 10 minutes, but stands out as it
sits well above the EHB (Fig.\,3), being
most likely in a post-EHB stage of evolution.

It took several years from the announcement of the first sdB
pulsator until it was realised that the same group of stars are subject
to longer period pulsations as well. Green et al.~(2003) published
the discovery of pulsations in V1093\,Her, with periods between one half
and two hours, and reported that as many as 75\% of sdB stars cooler
than 30\,kK display some level of pulsations at these periods.
The V1093\,Her stars span the EHB from
the coolest sdBs at around 24\,kK up to the domain of the
V361\,Hya stars (Fig.\,3).
The pulsations were identified with
high radial order $g$-modes by Green et al.~(2003), and their
amplitudes are very low, typically 0.2\,\% or less.
Although such low level pulsations are common in V1093\,Her stars,
and have been detected in at least 30, only a few have been studied
in detail due to the long time-base and high precision required
to detect and resolve their modes.

Even more recently, Schuh et al.~(2006) realised that a known
V361\,Hya star, DW\,Lyn, was actually
displaying the $g$-modes of a V1093\,Her star simultaneously with
$p$-modes of a V361\,Hya star. As noted by the authors, the four
V361\,Hya stars DW\,Lyn, V391\,Peg and Balloon 090100001 (BA09
in Fig.\,3 and hereafter), and KL\,UMa, form a compact group
closer to the domain of the V1093\,Her stars than the remaining
V361\,Hya stars. Hybrid DW\,Lyn type pulsations have now been
detected in both V391\,Peg and BA09 as well, but appear to be
absent in KL\,UMa. Intriguingly, KL\,UMa also stands out as the only
binary of the quartet (O'Toole et al.~2004).

A fourth type of pulsations was noted in the
He-sdB star LS\,IV--14$^\circ$116 by Ahmad \& Jeffery (2005).
They detected pulsations with amplitudes at the 1\% level and
periods around 15\,minutes. The atmospheric parameters reported
by the authors, \teff\,=\,32.5\,kK, \logg\,=\,5.4\,dex places
the star just above the EHB strip in the \teff--\logg\ diagram,
well surrounded by V361\,Hya stars. With a supersolar helium
abundance, log(He/H)\,=\,-0.6, this star represents a different
evolutionary state than the regular EHB pulsators. Up to now
this star remains unique, but since stars with the atmospheric
properties of LS\,IV--14$^\circ$116 are extremely rare it
is too early to tell whether pulsations in stars with similar
atmospheric parameters are common or rare.

A fifth and final class of pulsations in hot subdwarf stars
was discovered by Woudt et al.~(2006) in the hot sdO binary
J17006+0748, but this will not be discussed here as this star
is very far from the EHB region.

\subsection*{Driving the Beat}

Pulsations in sdB stars were predicted to occur by Charpinet et al.~(1996)
at about the same time as the first pulsators were discovered by
Kilkenny et al.~(1997).
The driving mechanism is due to an opacity bump caused by iron group
elements (Charpinet 1997). This mechanism is inefficient at solar metallicity,
but gravitational settling and radiative levitation can work together to
locally enhance metals in a driving zone in the envelope.
This $\kappa$ mechanism has been successfully invoked to explain both
the $p$-mode pulsations in V361\,Hya stars and the $g$-mode
pulsations in V1093\,Her stars (Fontaine et al.~2003).

While the first models by Fontaine et al.~(2003) could produce
unstable modes in the coolest sdB stars, there appeared to be a gap
between one island of instability on the cool end of the EHB and one
at the hot end. The pulsators at the hot end of the $g$-mode instability
region remained problematic, and particularly so
the hybrid DW\,Lyn type pulsators. Some relief to this problem was recently
provided by Jeffery \& Saio (2007), with the application of improved
opacity values from the OP project as well as the explicit consideration
of nickel in addition to iron. The new models are sufficient to bridge
the gap between the hot and cool EHB pulsators, and could also predict
an island of instability in the sdO domain, close to the observed location
of J17006+0748. However, these most recent calculations do not yet give
a perfect description of the observed picture. More unstable modes are
still found in models at the hot end of the EHB than on the cool end,
while observations indicate that pulsations are more common in cool
sdB stars. In fact, the problem is now more to explain why most EHB stars
on the hot end of the strip do not pulsate, than why they do.
Jeffery \& Saio (2007) speculated that the iron group element
enhancements, which build up due to a diffusion process over rather long
timescales, may be disrupted by the the atmospheric motions as pulsations
build up to some level. They note that since $p$-modes mostly involve
vertical motion, while $g$-modes are dominated by horizontal motion,
it is possible that $p$-modes are more effective at redistributing the
iron group elements out of the driving zone.

\subsection*{Levitation does the Trick}

Detailed models of the internal structure of the EHB stars are
critical for improving our understanding of the asteroseismic
properties. The simplest models with uniform metallicities are
not able to drive pulsations in these stars at all. Only with the
inclusion of an iron opacity bump was it possible to find unstable
modes (Charpinet 1996). 
However, the periods predicted by these so-called second generation
models have usually not been matched with observed periods to better than
about one percent, while the observed periods have a precision that
are an order of magnitude better.
Efforts to improve the atmospheric models to include
more of the various effects that can have significant impact on the
pulsation spectrum are ongoing.

Important progress was reported by Fontaine et al.~(2006), who
clearly demonstrated the importance of properly including
time-dependent diffusion calculations in order to predict pulsation
frequencies and mode stability. Starting from a uniform distribution
and solar iron abundance, they demonstrated that it takes several
hundred thousand years for radiative acceleration and gravitational 
settling to produce sufficient iron in the driving region to
create unstable modes. After about 1\,Myr there are no more changes
with respect to which modes are driven and not, but the pulsation
frequencies may still shift as the iron opacity bump builds up
further. After about 10\,Myr iron reaches equilibrium in these 
models, and no further changes are seen.
Since the time to onset of pulsations is just 1/1000 of the typical
EHB lifetime, this does not help resolve the issue as to why only
a fraction of the EHB stars pulsate.

\figureCoAstII{seismo_MMe}{seismo_MeG}
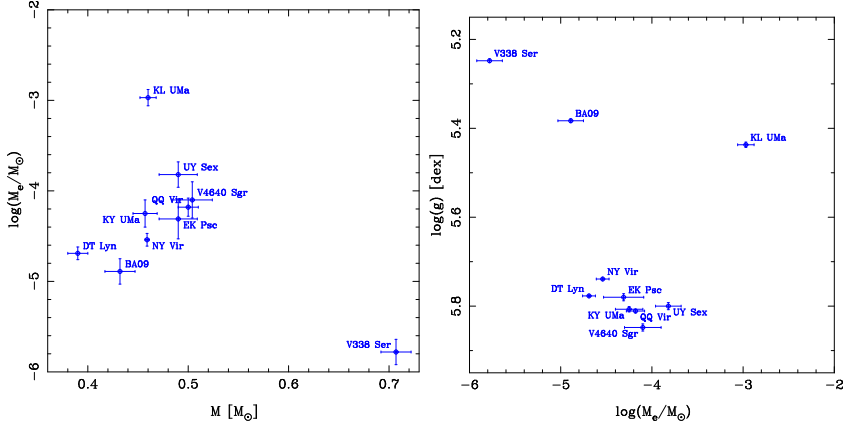
{Published asteroseismic solutions for ten V361\,Hya stars.
The left hand panel shows total mass plotted versus envelope mass,
and the right hand panel shows envelope mass versus surface gravity.
Note that only the optimal solution is shown even if the
papers discuss several possible ones.
}
{fig:seismo}{t}{clip,angle=0,width=0.48\textwidth}

Second generation models based on a pure hydrogen atmosphere on top
of a simple 'hard ball' core approximation, to which an
explicit iron abundance profile is added,
have been used to derive sensible asteroseismic quantities
for a number of V361\,Hya stars (e.g. Charpinet et al.~2007).
The adopted `forward' method basically consists of constructing a large
grid of models in the four dimensional parameter space spanned by
the fundamental model parameters, effective temperature \teff,
surface gravity \logg, total mass $M$, and envelope mass fraction $q$\/(H).
A minimisation procedure is then invoked to find the model that best
matches the observed periods.

To date, ten asteroseismic solutions computed with this method have been
published.
They were summarised in Randall et al.~(2007) for the
first seven, and in Fig.\,4 the new solutions by van Grootel et al.~(2008),
van Spaandonk et al.~(2008) and Charpinet et al.~(2008) (discussed below)
have been included.
A feature of the asteroseismic modelling is that \teff\ is rather
poorly constrained, and a better value can usually be
provided from spectroscopy. The surface gravity, total mass, and envelope
mass fraction, however, have very small associated errors in the
asteroseismic solutions, so we plot only these in Fig.\,4.
The distribution of masses is not as concentrated around 0.47\,M\sol\ as
most canonical evolutionary models have presumed,
but all points are well within the permitted ranges for synthetic
populations considered by Han et al.~(2002, 2003).
Except for two outliers, all the stars
appear to form a trend with envelope mass, $M_{\mathrm e}$,
increasing with total mass, $M$.
Although this feature has not been accounted for by evolutionary
calculations, it could occur as a natural consequence of a 
higher core mass requiring more energy to remove the envelope.
More disturbing is the lack of any clear trend in envelope mass
versus surface gravity, as is clearly demanded for canonical EHB models.
The scatter in the high gravity objects is easily explained by their spread
in mass, and KL\,UMa fits well with the expected $M_e$/\logg\ trend. But
the unusually low envelope masses for BA09 and V338\,Ser are hard to
explain, and may indicate that the adopted models are too simplified
to represent the seismic properties for these cases.

\subsection*{Ballooney}

The exceptional amplitude of the dominant period in BA09 has hinted
towards a radial nature, and this was finally confirmed by
Baran et al.~(2008) by combining evidence from multicolor photometry
and the radial velocity amplitude measured by Telting \& \O stensen (2006).

Van Grootel et al.~(2008) have successfully applied the forward
method to BA09, demonstrating some peculiarities in the model predictions.
Their optimal solution for the main mode, when using no constraints,
is $\ell$\,=\,2, which is not reconcilable with the spectroscopic data.
However, by imposing mode constraints from multicolor photometry they do
find asteroseismic solutions that agree with all observational data.
Curiously, the physical parameters for the constrained and unconstrained
fits are almost identical, even if the mode identification changes for
half the modes considered. The authors conclude
``{\em Our primary result is that the asteroseismic solution stands very
robust, whether or not external constraints on the values of the degree
$\ell$ are used.}'' This peculiarity arises from the high
mode density and the way the modes are distributed in period space.
But the deeper cause of the problem is the low precision with which the
second generation models predict pulsation periods. Models with
a more detailed internal structure are therefore urgently needed
in order to resolve this problem.
It is a concern that the envelope mass fraction
van Grootel et al.~(2008) find ($\log M_e/M_\ast$\,$\simeq$\,--\/4.9,
regardless of which modes are which)
is several hundred times lower than what
any EHB model would predict for such a low mass star at this position
in the \teff--\logg\ diagram.
With such a thin envelope all models put the star close to
the HeMS for a core helium burning star.
The authors' suggestion that the star is in a post-EHB stage of
evolution is beyond canonical theory, as only models with substantial
hydrogen envelopes evolve to lower gravities before moving off to
the sdO domain as the core starts to contract (see Fig.\,1).

The envelope mass discrepancy is even more severe in the asteroseismic
results for V338\,Ser (van Spaandonk et al.~2008), whose best fitting
model (number 4 of 5 presented) has an envelope mass fraction
$\log M_e/M_\ast$\,=\,--5.78! While the authors seem to prefer an
even more extreme value of --6.22 for a model with a slightly poorer
fit, in order to obtain a mass of 0.561\,M\sol\ rather than the unusually
high mass of 0.707\,M\sol, the high mass solution might be the most
interesting.
For if the exceptionally high mass is real, evolutionary calculations would
place V338\,Ser in a core helium burning stage, not in a post-EHB stage
as would be the case if its mass was around 0.5\,M\sol. But as for BA09,
evolutionary models demand an envelope mass fraction more than
1\,000 times higher than found by van Spaandonk et al.~(2008),
in order to find V338\,Ser at the observed \logg.

With a recent update of the forward modelling code, Charpinet et al.~(2008)
have produced a very convincing model for the eclipsing binary
system NY\,Vir.
This star has been particularly challenging since it is rapidly rotating,
due to being in a tidally locked orbit with the close M-dwarf companion.
The rotational splitting of modes with different
$m$ produces a particularly rich pulsation spectrum.
Charpinet et al.~(2008) use asteroseismology to discriminate between
three solutions from the binary orbit published by
Vu\v{c}kovi\'c et al.~(2007), and finds that the intermediate model
with a mass of 0.47\,M\sol\ is clearly favored. This solution is
also the only one consistent with the \logg\ from the BG survey (Fig.\,3).

Most recently, Telting et al.~(2008) have presented the first
study of line-profile variations in these stars
based on high-resolution spectroscopy.
Line profile variations of metal lines is a technique to directly
determine the spherical harmonic order numbers $\ell, m$ of a pulsation
mode, which is well established for various MS pulsators.
To invoke this technique on the faint EHB stars requires substantial
investments in terms of telescope time, which has prohibited its
use up to now. With the preliminary results
on the high amplitude pulsator BA09,
Telting et al.~(2008) clearly demonstrate
that the $\ell$ of the main mode must be either 0 or 1. Again, the
observational accuracy has advanced ahead of the theoretical models,
as the standard modelling of such line profile variations are insufficient
to properly account for the complex effects on the line profiles in the
high temperature and gravity domain of the EHB stars.

A final advancement that demonstrates with particular clarity the
direction in which asteroseismology of EHB stars needs to move in order
to progress, is the work of Hu et al.~(2008).
The authors have taken real evolutionary models that have been evolved
from the ZAMS, through flashless
helium ignition on the RGB, to the EHB by peeling
of the envelope, and proceeded to compute the pulsation properties of
these stars for a number of different configurations. By comparing
these models with the classical post-flash models with similar
surface parameters, they clearly show that the differences
in internal structure from the two evolutionary paths
produce significant differences in both the predicted pulsation periods,
and with respect to which modes are excited or damped!
Two of the flashless models of Hu et al.~(2008) are plotted in Fig.\,5
together with a classical post-flash EHB model. From an evolutionary
population synthesis point of view it is interesting to note that the
post-EHB tracks are substantially different. After the core has exhausted
its helium and the star moves off the TAEHB the star briefly burns the
remaining envelope hydrogen, before contracting and cooling down during a
helium shell burning phase, which lasts much longer than in canonical
models (up to 20\,Myrs). This is long enough to produce a slight clustering
of objects below the HeMS at temperatures between 45 and 50\,kK, just
where a substantial cluster of He-sdO stars are observed.

\figureCoAst{BG_TGHe+HH}{
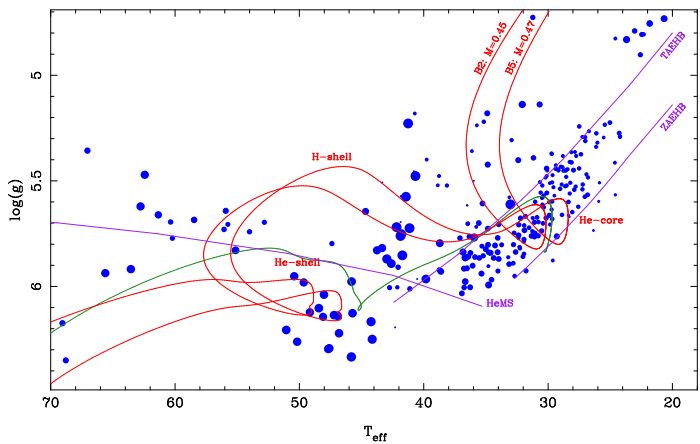
Same as Fig.\,1, but with the evolutionary tracks for flashless post-RGB
evolution from Hu et al.~(2008) overplotted.
The tracks evolve rapidly from the top of the plot
to the ZAEHB as the envelope settles down on the helium burning core,
and the stars spends most of their time (200\,Myr) in the upward part of
the loop at constant \teff. After core helium exhaustion the star
rapidly heats up, but then turns and moves back again for
a several Myr long shell helium burning phase below the HeMS.}
{fig:TG_HH}{t}{clip,angle=0,width=0.8\textwidth}

\section*{Conclusions}

Asteroseismology of EHB stars is a field in rapid progress, with
exceptional challenges due to their complex formation paths.
Much progress has been made on evolutionary models, but much remains
to be done, particularly with respect to the formation of single EHB stars.
Better models are also needed to reproduce the effects of the
helium flash stage on the envelope composition, in order to reproduce
the internal structure of EHB stars at the accuracy achieved by
observational asteroseismology. 
Only when these advances are in place can asteroseismology reliably
test the different formation scenarios.

The low amplitudes and long periods make it very difficult to establish
detailed pulsation spectra for V1093\,Her stars,
and even when it can be done the high mode density makes
it difficult to assign modes to the observed frequencies.
But $g$-modes are particularly interesting because they probe deep into
the stellar interior. This is a significant challenge for the future
due to the long time-base required to reliably determine the longer
pulsation periods in these stars. The upcoming Kepler satellite mission
(Christensen-Dalsgaard et al. 2007)
provides an excellent opportunity if pulsators can be found within its
field of view.

\acknowledgments{\small
Special thanks to E.M.\,Green
for kindly providing the detailed results
from her 2008 article, which made it possible to reproduce the
figures from that article together with evolutionary tracks.
Without these data, Fig.\,1, 2, 3, and 5 would have been a lot
less informative.
The author is supported by the Research Council of the University of Leuven
under grant GOA/2008/04
and by the EU FP6 Coordination Action HELAS.}

\setlength\smallskipamount{0.5 pt}
\References{
\rfr Ahmad, A., \& Jeffery, C.S., 2005, \aap, 437, L51
\rfr Baran, A, Pigulski, A, \& O'Toole, S.J., 2008, \mnras, 385, 255
\rfr Charpinet, S., Fontaine, G., Brassard, P., et~al.\ 1997, \apj, 483, L123
\rfr Charpinet, S., Fontaine, G., Brassard, P., \& Dorman, B., 1996,
	\apj, 471, L103
\rfr Charpinet, S., Fontaine, G., Brassard, P., et~al.\ 2007, CoAst, 150, 241
\rfr Charpinet, S., van Grootel, V., Reese, D., et al.~\ 2008, \aap, 489, 377
\rfr Christensen-Dalsgaard, J., Arentoft, T., Brown, T.M., et~al.\ 2007,
	CoAst, 150, 350
\rfr D'Cruz, N.L., Dorman, B., Rood, R.T., \& O'Connell, R.W., 1996,
	\apj, 466, 359
\rfr Driebe, T., Bl\"ocker, T., Sch\"onberner, D., \& Herwig, F., 1999,
	\aap, 350, 89
\rfr Edelmann, H., Heber, U., Hagen, H.-J., Lemke, M., et al.\ 2003
	\aap, 400, 939
\rfr Fontaine, G., Brassard, P., Charpinet, S., et al.\ 2003, \apj, 597, 518
\rfr Fontaine, G., Brassard, P., Charpinet, S., \& Chayer, P. 2006,
	Mem.~S.A.It., 77, 49
\rfr Green, E.M., Fontaine, G., Reed, M.D., et~al.\ 2003, \apj, 583, L31
\rfr Green, E.M., Fontaine, G., Hyde, E.A., et~al.\ 2008, 
	ASP Conf.~Ser., 392, 75
\rfr Green, R.F., Schmidt, M., \& Liebert, J., 1986, ApJSS, 61, 305
\rfr Greenstein, J.L., \& Sargent, A.I., 1974, ApJSS, 28, 157
\rfr van Grootel, V., Charpinet, S., Fontaine, G., et~al.\ 2008, \aap, 488, 685
\rfr Han, Z., Tout, C.A., Eggleton, P.P, 2000, \mnras, 319, 215
\rfr Han, Z, Podsiadlowski, Ph., Maxted, P.F.L, et~al.\ 2002,
	\mnras, 336, 449
\rfr Han, Z, Podsiadlowski, Ph., Maxted, P.F.L, \& Marsh, T.R., 2003,
	\mnras, 341, 669
\rfr Heber, U., 1986, \aap, 155, 33
\rfr Heber, U., Edelmann, H., Lisker, T., Napiwotzki, R., 2003, \aap, 411, L477
\rfr Heber, U., 2009, Annual Review of Astronomy and Astrophysics, 47, submitted
\rfr Humason, M.L., \& Zwicky, F., 1947, \apj, 117, 313
\rfr Iben Jr, I., \apj, 1976, 208, 165
\rfr Jeffery, C.S., \& Saio, H., 2007, \mnras, 378, 379
\rfr Kawaler, S.D., \& Hostler, S.R., 2005, \apj, 621, 432
\rfr Kilkenny, D., Koen, C., O'Donoghue, D., \& Stobie, R.S., 1997,
	\mnras, 285, 640
\rfr Lisker, T., Heber, U., Napiwotzki, R., et~al.\ 2005, \aap, 430, 223
\rfr Maxted, P.F.L, Heber, U., Marsh, T.R., \& North, R.C., 2001,
	\mnras, 326, 1391
\rfr Miller Bertolami, M.M., Althaus, L.G., Unglaub, K., \& Weiss, A., 2008,
	\aap, 491, 253
\rfr Moehler, S., Richtler, T., de Boer, K.S., et~al.\ 1990, A\&ASS, 86, 53
\rfr Napiwotzki, R., Karl, C.A., Lisker, T., et~al.\ 2004, AP\&SS, 291, 321
\rfr Nelemans, G., \& Tauris, T.M., 1998, \aap, 335, 85
\rfr Newell, E.B., 1973, ApJSS, 26, 37
\rfr Oreiro, R., \O stensen, R.H., Green, E.M., \& Geier, S., 2009,
	\aap, in press.
\rfr \O stensen, R.H., 2006, Baltic Astronomy, 15, 85
\rfr O'Toole, S.J., Heber, U., Benjamin, R.A., 2004, \aap, 422, 1053
\rfr Paczy\'nski, B., 1971, Acta Astronomica, 21, 1
\rfr Podsiadlowski, Ph., Han, Z., Lynas-Gray, A.E., \& Brown, D., 2008,
	ASP Conf.~Ser., 392, 15
\rfr Politano, M., Taam, R.E., van der Sluys, M., Willems, B., 2008,
	\apj, 687, L99
\rfr Saio, H., \& Jeffery, C.S., 2000, \mnras, 313, 671
\rfr Schuh, S., Huber, J., Dreizler, S., Heber, U., et~al.\ 2006,
	\aap, 445, L31
\rfr Silvotti, R., Schuh, S., Janulis, R., Solheim, J.-E., et~al.\ 2007,
	Nature, 449, 189
\rfr Soker, N., 1998, AJ, 116, 1308
\rfr van Spaandonk, L., Fontaine, G., Brassard, P., \& Aerts, C., 2008,
	ASP Conf.~Ser., 392, 387
\rfr Stoughton, C., Lupton, R.H., Bernardi, M., et~al.\ 2002, AJ, 123, 485
\rfr Stroeer, A., Heber, U., Lisker, T., et~al.\ 2007, \aap, 462, 269
\rfr Sweigart, A.V., 1997,
	in:{\em ``The Third Conference on Faint Blue Stars''},
	Philip, Liebert, Saffer \&\ Hayes (Eds), L.~Davis Press, p.3
	(arXiv:astro-ph/9708164)
\rfr Telting, J.H., \& \O stensen, R.H., 2006, \aap, 450, 1149
\rfr Telting, J.H., Geier, S., \O stensen, R.H., Heber, U., et~al.\ 2008,
	\aap, in press.
\rfr Vu\v{c}kovi\'c, M., Aerts, C., {\O}stensen, R., et~al.\ 2007,
	\aap, 471, 605
\rfr Webbink, R.F., 1984, \apj, 277, 355
\rfr Woudt, P.A., Kilkenny, D., Zietsman, E., et~al.\ 2006, \mnras, 371, 1497
}

\end{document}